\documentclass[aps,prl,amsmath,twocolumn,amssymb,titlepage,superscriptaddress,showpacs]{revtex4-1}
\usepackage{graphicx}
\usepackage{setspace}

\usepackage{color}
\usepackage{ulem}

\usepackage{amsmath,amsthm,amssymb}
\usepackage{mathrsfs}

\newcommand{\ave}[1]{\langle #1\rangle}

\newcommand{\I}{\mathrm{i}}

\begin{document}
\title{Nonlinear spectroscopy of collective modes in excitonic insulator}
\author{Denis Gole\v{z}}
\affiliation{Center for Computational Quantum Physics, Flatiron Institute, 162 Fifth Avenue, New York, NY 10010, USA}
\author{Zhiyuan Sun}
\affiliation{Department of Physics, Columbia University, 538 West 120th Street, New York, NY 10027}
\author{Yuta Murakami}
\affiliation{Department of Physics, Tokyo Institute of Technology, Meguro, Tokyo 152-8551, Japan}
\author{Antoine Georges}
\affiliation{Center for Computational Quantum Physics, Flatiron Institute, 162 Fifth Avenue, New York, NY 10010, USA}
\affiliation{Department of Quantum Matter Physics, University of Geneva, 1211 Geneva 4, Switzerland}
\affiliation{CPHT, CNRS, Ecole Polytechnique, IP Paris, F-91128 Palaiseau, France}
\affiliation{Collège de France, 11 place Marcelin Berthelot, 75005 Paris, France}
\author{Andrew J. Millis}
\affiliation{Center for Computational Quantum Physics, Flatiron Institute, 162 Fifth Avenue, New York, NY 10010, USA}
\affiliation{Department of Physics, Columbia University, 538 West 120th Street, New York, NY 10027}

\date{\today}

\begin{abstract}
The nonlinear optical response of an excitonic insulator coupled to lattice degrees of freedom is shown to depend in strong and characteristic ways on whether the insulating behavior originates primarily from electron-electron or electron-lattice interactions. Linear response optical signatures of the massive phase mode and the amplitude~(Higgs) mode are identified. Upon nonlinear excitation resonant to the phase mode, a new in-gap mode at twice the phase mode frequency is induced, leading to a huge second harmonic response. Excitation of in-gap phonon modes leads to different and much smaller effects. A Landau-Ginzburg theory analysis explain these different behavior and reveals that a parametric resonance of the strongly excited phase mode is the origin of the photo-induced mode in the electron-dominant case.  The difference in the nonlinear optical response serve as a measure of the dominant mechanism of the ordered phase.
\end{abstract}

\maketitle

{The excitonic insulator is a quantum state of matter characterized by the spontaneous formation of an interband hybridization driven by electron-electron interactions. While the state was theoretically predicted long ago~\cite{jerome1967,keldysh1968,kohn1967}, the identification of material systems that exhibit the phase has lagged.  Photoluminescence measurements \cite{butov2002} followed by the demonstration of dissipationless Coulomb drag  and interlayer coherence \cite{eisenstein2004bose,wang2019evidence,nandi2012exciton} have  unambiguously identified the excitonic phase in  specifically designed semiconductor double-layer heterostructure systems in the quantized Hall regime of high magnetic fields and cryogenic temperatures. More recently,  several transition metal chalcogenide materials including  Ta$_2$NiSe$_5$~(TNS)~\cite{sunshine1985,wakisaka2009excitonic,kaneko2012,seki2014excitonic} and TiSe$_2$~\cite{monney2009,monney2012,kaneko2018exciton} that exhibit gap opening transitions as the temperature is lowered have been proposed as excitonic insulator candidates.   However, in these materials the gap opening transition is accompanied by a structural distortion, which can also open a gap, while on the dynamical level both  electron-electron \cite{seki2014excitonic,mazza2020} and electron-phonon ~\cite{kaneko2013,subedi2020}  interactions appear to be strong.  The question of the dominant physics in these materials is thus the subject of intense current interest~\cite{mor2017,okazaki2018,tang2020,baldini2020,kim2020,andrich2020b}.

Collective excitations are important signatures of  quantum coherent states and their observation can provide key insight. A quantum state characterized by a spontaneously broken  $U(1)$ symmetry typically exhibits two characteristic modes: a gapless phase~(Goldstone) mode  and a gapped amplitude~(Higgs) mode. The insights provided by collective mode spectroscopy of superconductors~\cite{Anderson1963plasmons,endres2011,zwerger2004anomalous,littlewood1982,pekker2015,matsunaga2014light,tsuji2015theory,matsunaga2017,tsuji2015theory,yusupov2010,Basov2017,sun2020} motivate investigations of the collective modes of excitonic insulators. 

Previous work has shown that excitonic order leads to a nontrivial change in the linear response to an electro-magnetic field~\cite{murakami2020}. In particular, the original theoretical proposals~\cite{jerome1967,keldysh1968,kohn1967} discussed an excitonic insulator state  that spontaneously breaks  an internal $U(1)$ symmetry related to the separate conservation of charge in the two bands. This suggests that the observation of a gapless phase mode would be a ``smoking gun" indication of excitonic order. However, ``real materials" effects complicate the issues. Interband hybridization is generically nonvanishing except at certain high-symmetry k-points, so the $U(1)$ symmetry is  reduced to a discrete (typically $Z_2$) symmetry~\cite{mazza2020,watson2020}, implying a gap in the phase mode~\cite{zenker2014}. Further, the excitonic order parameter couples linearly to certain classes of lattice distortions~\cite{kaneko2013}, so these lattice distortions will generically appear at the excitonic phase transition and will further increase the gap in the phase mode. Additionally, the electron lattice coupling means that excitonic modes are coupled to phonon modes, so excitations are generically of mixed electronic and lattice character~\cite{kaneko2013} and it is not straightforward to disentangle the collective modes from phononic response in standard linear-response experiments. Going beyond linear response, recent experiments including optical pump-induced anomalous transport~\cite{andrich2020} and phononic response~\cite{werdehausen2018} have reported very interesting indirect signatures of a collective response, but difficulties in disentangling electronic and phononic effects and the lack of a direct signature means that the issue is not yet settled.

In this work, we show that the nonlinear response of the material to strong electromagnetic excitation at subgap frequencies provides a powerful method of distinguishing electronic and phononic mechanisms for the gap opening transition.

We focus on the simplest model that contains the relevant physis, namely a two-band model of spinless fermions in one dimension coupled to a single dispersionless phonon mode and to an applied electric field, described by the Hamiltonian
\begin{equation} \begin{split}
H=&\sum_{k,\alpha \in \{0,1\}}
(\epsilon_{k- A,\alpha}-\mu)c_{k,\alpha}^\dagger c_{k,\alpha} + V \sum_{i} n_{i,0} n_{i,1}\\
&+\sum_i \left[\sqrt{\lambda}\hat{X}_i - E(t)  D \right] (c_{i,1}^{\dagger} c_{i,0}+c_{i,0}^{\dagger} c_{i,1}) \\
&+\sum_i \frac{1}{2}\left(\hat{X}_i^2+\frac{1}{\omega_0^2}\dot{\hat{X_i}}^2\right) + \text{h.c.},
\label{Eq:Ham}
\end{split}
\end{equation}
Here $c^\dagger_{k\alpha}$ creates an electron of momentum $k$ in band $\alpha=0,1$. The bare electron dispersion $\epsilon_{k, \alpha}=-2J_\alpha\cos(k) + (-1)^{\alpha}\Delta\epsilon$ involves a band offset $\Delta\epsilon$ and hopping integrals $J_0=-J_1=-1$ chosen so that the bands disperse in opposite directions and are inverted at the $\Gamma$ ($k=0$) momentum point as shown by the dashed lines in panel (a) of Fig.~\ref{Fig:optics_beta}. The chemical potential $\mu$ is chosen such that the system is at half filling.  $X_i$  is the atomic displacement on site $i$  normalized so that the spring constant is unity, and $\lambda$ is the  electron-phonon coupling and we have assumed the phonons are dispersionless with mode frequency $\omega_0$.    Following  Ref.~\onlinecite{li2020,golez2019,golevz2019dynamics} the electric field is represented  by the minimally coupled Coulomb-gauge vector potential $A$ for the intra band terms and by a dipolar coupling to the electric field $E=-\partial_tA$ for the interband terms with dipole moment $D=1$, chosen here to be real and constant for simplicity.

In the absence of electron-phonon interaction ($\lambda=0$), the model separately conserves charge in the two bands and accordingly has an internal  $U(1)$ symmetry; for $\lambda\neq 0$  the symmetry is broken. The symmetry breaking is analogous to the hybridization-induced  symmetry breaking discussed in  Ref.~\onlinecite{mazza2020,watson2020}  so we do not include the latter effect explicitly. 

The electron-electron interaction is treated in the time-dependent Hartree-Fock approximation; see Ref.~\onlinecite{murakami2017Photo,tanabe2018,tanaka2018photoinduced} for details. The interorbital interaction is factorized as $V\sum_in_{i,0}n_{i,1}\rightarrow -V(\phi c^\dagger_{i,1}c_{i,0}+H.c.)$ where a nonzero value of the order parameter  $\phi(t)=\ave{c_{i,0}^{\dagger} c_{i,1}}$ signals the presence of interband~(excitonic) coupling. Moreover, we include the diagonal factorization in the orbital indices as $V\sum_i n_{i,0}n_{i,1}\rightarrow V(n_{i,0}\langle n_{i,1}\rangle+H.c.)$ and the mean  lattice distortion $X(t)=\left<\hat{X}\right>$~\cite{murakami2017Photo,golevz2019dynamics,cilento2018,lantz2017}.

The equilibrium phase diagram following from this approximation  is discussed in Ref.~\onlinecite{kaneko2013}. For nonzero $\lambda$ or $V$, the state has both $X$ and $\phi$ $\neq 0$, except in the special case $\lambda=0$ where $X$ remains zero.  A nonzero $\phi$ or $X$ hybridizes the two bands, opening a gap in the spectrum as shown by the solid line in Fig.~\ref{Fig:optics_beta}. The gap magnitude is determined by the parameter $V^\star=V+2\lambda$. We  distinguish ``primarily electronic'' and ``primarily lattice'' situations according to whether $V$ or $\lambda$ gives the dominant contribution to $V^\star$.

\begin{figure}[t]
\includegraphics[width=1.0\linewidth]{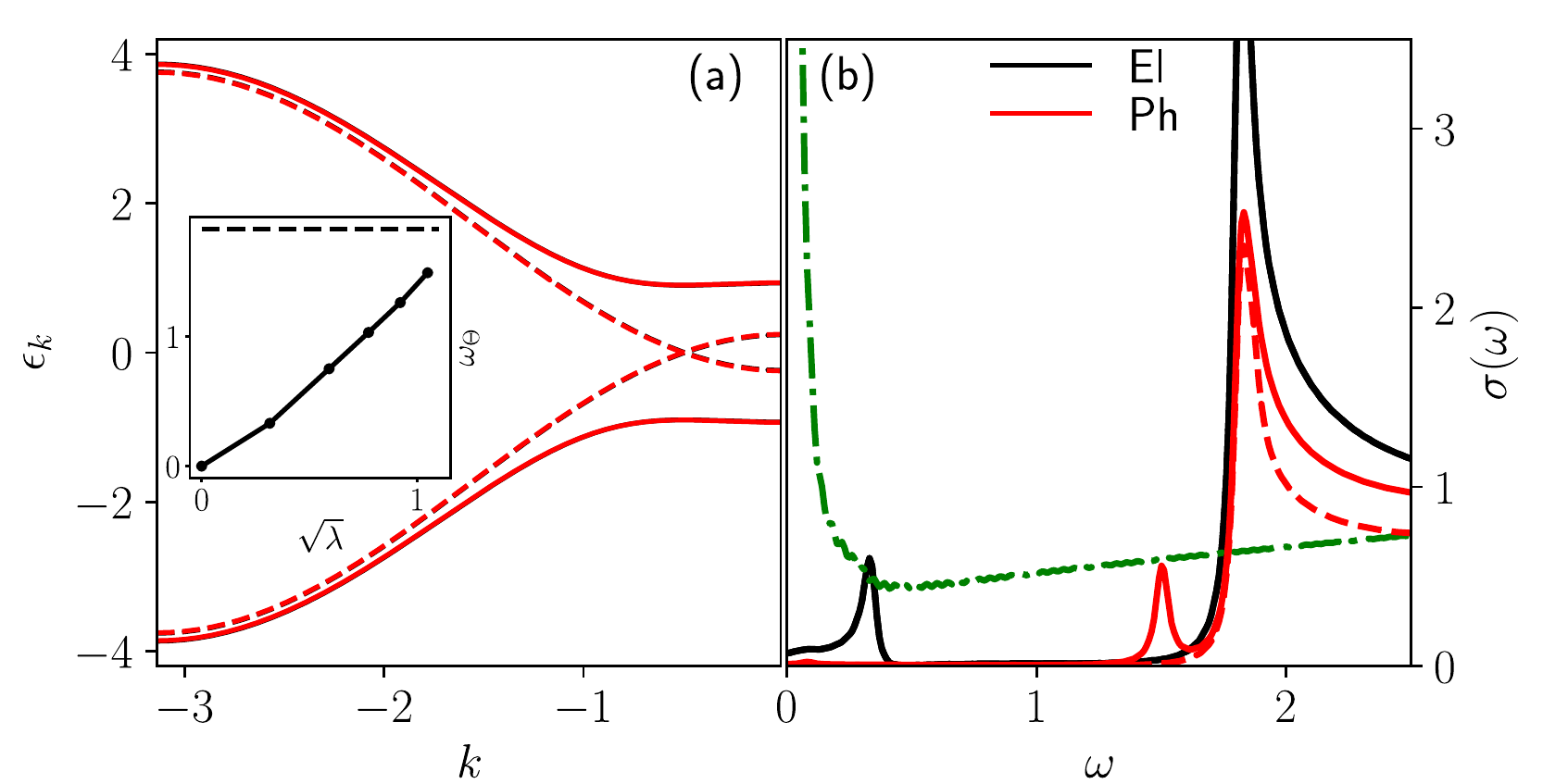}
\caption{(a)~Single particle dispersion for symmetry unbroken (dashed line) and symmetry broken (solid lines) states. The inset shows the evolution of the phase mode frequency $\omega_{\Theta}$ with the electron phonon coupling $\lambda$ and the horizontal dashed line represents the single-particle gap.~(b) Linear optical conductivity of equilibrium  insulating states in   the ``primarily lattice''~(red solid line, parameters (V=1.0,~$\lambda$=1.1,~$\Delta\epsilon=-1.13$, $\omega_0=0.1$) and ``primarily electronic'' case~(black solid line, parameters (V=3.0,~$\lambda$=0.1,~$\Delta\epsilon=-2.4$, $\omega_0=0.1$). The quasiparticle gap at $2\Delta\approx 1.8$ is visible as a large amplitude peak; the small in-gap peaks arise from excitation of the phase mode. The dash-dotted line shows the optical response in the metallic state.  The dashed lines show the linear response  optical conductivity of the two insulating states calculated  without the vertex correction. All insulating computations are performed at $T=0$.}
\label{Fig:optics_beta}
\end{figure}

We now compare linear and nonlinear optical spectra in ``primarily electronic'' and ``primarily lattice'' cases with parameters chosen so that  characteristic cases at zero temperatures and in the BCS regime. We have adjusted model parameters such that the single particle gaps $\Delta=\phi+\sqrt{\lambda}X=1.8$ are the same in both cases. The solid lines in Fig.~\ref{Fig:optics_beta}(b) present the equilibrium linear response conductivities for the two insulating states. The response in both cases displays a large peak associated with excitation of carriers across the insulating gap and a smaller in-gap peak.  The in-gap feature is a direct signature of the collective response, which consists of phase and amplitude modes~\cite{murakami2020}. In this model the phase mode is gapless at $\lambda=0$ and its minimum frequency is very sensitive to strength of the $U(1)$ symmetry breaking terms; as the magnitude of the symmetry breaking (here expressed  by $\lambda$) increases the phase mode frequency  rapidly approaches twice the single-particle gap energy (see the inset in Fig.~\ref{Fig:optics_beta}(a))~\cite{murakami2020}. The phase mode appears in the optical spectrum due to the nonzero dipolar moment $D$ between the conducting and the valence band. The frequency $\omega\approx 1.8$ of the larger amplitude conductivity feature corresponds to the minimum energy required to excite a particle-hole pair across the insulating gap. However, as can be seen from the difference between the solid and dashed lines in Fig.~\ref{Fig:optics_beta}(b), collective modes~(mathematically expressed as vertex corrections~\cite{schrieffer2018theory}) significantly affect the amplitude of the response, with the effect being much larger in the ``primarily electronic'' case than in the ``primarily lattice'' case. 
\begin{figure}[t]
\includegraphics[width=1.0\linewidth]{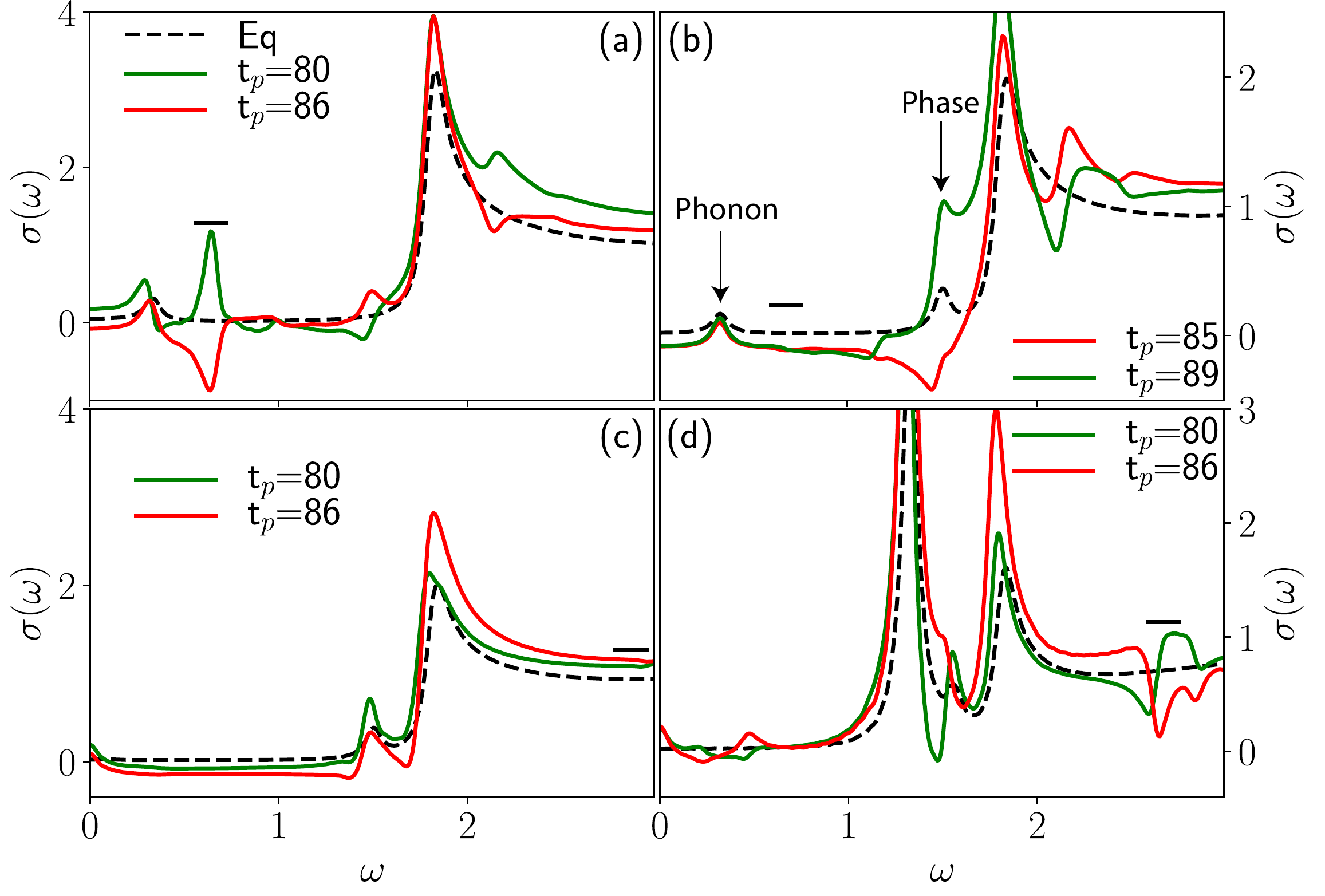}
\caption{Nonequilibrium optical conductivity defined as linear response to a delta function electric field pulse applied at delay times $t_p$ after a pump pulse. Nonequilibrium response of: (a)~``primarily electronic'' driven insulator with the central frequency tuned to phase mode frequency $\Omega$=0.4, (b)~``primarily lattice'' driven insulator with phonon and pump frequency set to $\omega_0$=$\Omega$=0.4, (c)~``primary lattice'' driven insulator with phonon frequency $\omega_{0}=0.1$  for excitation resonant to massive phase mode $\Omega$=1.4, (d) ``primarily lattice'' driven insulator with phonon frequency and pump frequency set to the gap edge $\omega_{0}$=$\Omega$=1.4. The shapshots are choosen such to represent the extreme changes in the optical conductivity. In all cases the pump strength is set to E$_0$=0.1. The heavy horizontal line is at $2\omega$ and the width is 0.15.}
\label{Fig:phase}
\end{figure}

The weakness of enhancement of the amplitude mode in ``primarily lattice'' driven case takes place as the phonon contribution to the vertex corrections only affects the collective motion at low energies comparable to the phonon frequency~\cite{murakami2020}. To experimentally assess the strength of this contribution to the conductivity we consider the f-sum rule which implies that the  optical integral is essentially the same in the gapped low-T and ungapped high-T states, with the spectral weight in the low frequency ``Drude" conductivity observed in the high temperature state simply transferred to the above gap absorption.  We find  that 80\% of the Drude weight~(integral up to $\omega=2.5$) appears at the near gap regions in the ``primarily electronic'' driven case, but only $\sim 50\%$  appears there in the lattice dominated case, with the remainder in each case distributed among higher energies.   Interestingly, a very strong and sharp gap edge feature, seemingly incompatible with a simple quasiparticle picture, has been observed in spectroscopic ellipsometry measurements of Ta$_2$NiSe$_5$, see Ref.~\onlinecite{larkin2017}. The feature was previously  attributed to either an exciton-phonon bound state~\cite{larkin2017} or strong electron-hole fluctuations~\cite{sugimoto2018strong}. The results presented here suggest that the observations may be understood as a collective response.

Our analysis shows that the equilibrium linear response conductivity is sensitive to phase mode fluctations and is therefore a very sensitive probe of a relative electronic and lattice contribution to the gap. However, in ``real materials''  the high transition temperature means that the Drude peak will be very broad, making the analysis of the near gap spectral weight more difficult to perform. Moreover, the presence of optically active phonons, which may also be coupled to electrons, means that the interpretation of subgap features is not unambiguous.

We now we will show that nonlinear spectroscopy can largely resolve these ambiguities. We model the nonlinear~(pump-probe) spectroscopy by exciting the system with a pump field
\begin{equation}	
	E(t)=E_0 \sin(\Omega t)e^{-4.6t^2/t_0^2},
	\label{Eq:pump}
\end{equation}
characterized by a center frequency $\Omega$ and the width $t_0=2\pi/\omega$ leading to an oscillation cycle. The system is then characterized by computing the response to a probe field taken to be a delta-pulse applied at a later time $t_p$.

The panel (a) of Fig.~\ref{Fig:phase} presents the nonequilibrium conductivity computed at two delay times after application of a pump with center frequency  $\Omega$ resonant with the phase mode frequency $\omega=0.4$. The pump induces coherent phase oscillations which have profound consequences for the optical response. A new in-gap state at twice  the phase mode frequency appears and its amplitude  oscillates in time with  a magnitude exceeding the amplitude of the equilibrium phase mode feature. Panel (b) of Fig.~\ref{Fig:phase} shows the nonlinear optical response of a ``primarily lattice'' system with phonon tuned to $\omega_{0}=0.4$. The resonant pump pulse induces coherent phonon oscillations with a maximum phonon distortion  comparable to the phase mode distortion. However, there is negligible response at twice the phonon mode frequency. Instead, the optical signal is modified at the energies of the phase mode and the gap edge transition. 

Second row in Fig.~\ref{Fig:phase} presents an equivalent analysis for the ``primarily lattice'' cases. Panel (c) shows  that a low energy phonon with $\omega_0=0.1$ and the pump excitation resonant to the massive phase mode~($\omega=1.4$) induces phase oscillations, however, the second harmonic in the optical response is pushed into the single-particle continuum where it is overdamped and the corresponding weight is very weak. Panel (d) shows the case where the phonon frequency is comparable to the gap edge~(V=1.0,~$\lambda$=1.1, $\omega_0=1.4$) and the phase mode is weak. In this situation, the coherent phonon oscillations induce optical changes at twice the phonon frequency, see the horizontal line in Fig.~\ref{Fig:phase}(d), as well as significant oscillations at the gap edge. In summary, pumping a low frequency phase mode produces a strong response at twice the mode frequency, whereas pumping a low frequency  phonon produces effects at the gap edge. By contrast, if the phase mode energy is close to the quasiparticle gap energy then pumping it produces small changes, whereas a near gap edge phonon would produce both strong changes in the gap edge absorption and a second harmonic response.

Now, we want to build an intuitive understanding of the distinct optical response for the resonantly excited lattice and phase degrees of freedom. To this end, we start with the Landau-Ginzburg theory for the order parameter dynamics $\phi=|\phi| \exp^{-\I \theta}$~\cite{Note1}. The electron-phonon coupling provides a finite mass for the phase mode and its equation of motion is described by a driven harmonic oscillator 
\begin{equation}
	\ddot \theta(t) = -\left[\omega_{\theta}^2+\gamma_{\theta} E(t)\right]\theta(t) + \alpha E(t).
	\label{Eq:phi}
\end{equation}
For simplicity, we have assumed that the absolute value of the order parameter $|\phi|$ and the phonon distortion $X$ are finite but fixed. For weak electron-phonon interaction ($V\approx V^*$), the gap of the massive phase mode  is given by $\omega_{\theta}=2\Delta \sqrt{\frac{1}{\nu V} \frac{\lambda}{V^\ast}},$ where we have introduces the  density of states $\nu,$ and it is rapidly growing with the increased electron-phonon interaction $\lambda$, see Fig.~1(b). The linear coupling to the external field is given by  $\alpha$ and the strength of the nonlinear coupling is $\gamma_{\theta}=D \Delta (V^\ast/V) \left(\frac{2}{\nu V^\ast} -1 +\frac{V}{V^\ast}\right).$  The nonlinear coupling in the equation of motion $\gamma_{\theta}E\theta$ leads to the \textit{the parametric resonance}. First, we assume that the initial pump pulse induces coherent phase oscillations  with the amplitude $\theta_0$ and the frequency $\omega_\theta$. The corresponding photo-induced change in the optical response due to the parametric resonance results in
\begin{align}
\Delta \sigma(\omega)=\nu(\frac{V}{V^*})  \gamma^2_{\theta}  \theta_0^2 \frac{i\omega}{\omega^2-4\omega_{\theta}^2}.
\label{Eq:ParResOptics}
\end{align}
The parametric resonance exhibits a strictly positive peak structure at twice of the massive phase mode frequency $\omega=2\omega_{\theta},$ which scales quadratically with the amplitude of the phase oscillations $\theta_0^2.$ 

The effective action for the lattice displacement leads to an equation of motion equivalent to Eq.~\ref{Eq:phi}, where the oscillation frequency is  given by the phonon frequency $\omega_0$~(neglecting the renormalization effects). The parametric resonance response will depend on the size of the nonlinear coupling $\gamma_{\text{ph}}\approx\frac{\nu D\Delta}{4}\lambda\frac{\omega_0^2}{\Delta^2}$ and the maximum phonon distortion~(expressed in dimensionless units) induced by the pump pulse $u_0=\sqrt{\nu} \max[X(t)-X(0)].$ The relative weight of the lattice and phase mode contribution to the parametric resonance response in the optical conductivity is given by
\begin{equation}
	\frac{\Delta\sigma_{\text{ph}}}{\Delta\sigma_{\theta}} \propto (\lambda\nu)^2 (\frac{\omega_0}{\Delta})^4
\end{equation}
where the proportionality includes the dimensionless coupling constants of the order of 1~\cite{Note1}. The parametric resonance response is strongly suppressed for $\omega_0\ll\Delta,$ which is the relevant situation in the candidate excitonic materials, like Ta$_2$NiSe$_5$~\cite{larkin2018,mor2018} and 1T-TiSe$_2$~\cite{weber2011}.  The distinct optical response for the resonantly driven in-gap modes can be traced to a much smaller nonlinear coupling of the electromagnetic field to the low-energy phononic degrees of freedom in comparison to the massive phase mode. However, if the coherently driven phononic mode is comparable to the gap size similar signature at twice  the base frequency can emerge, which is consistent with an example in Fig.~\ref{Fig:phase}(d).

\begin{figure}[t]
\includegraphics[width=1.0\linewidth]{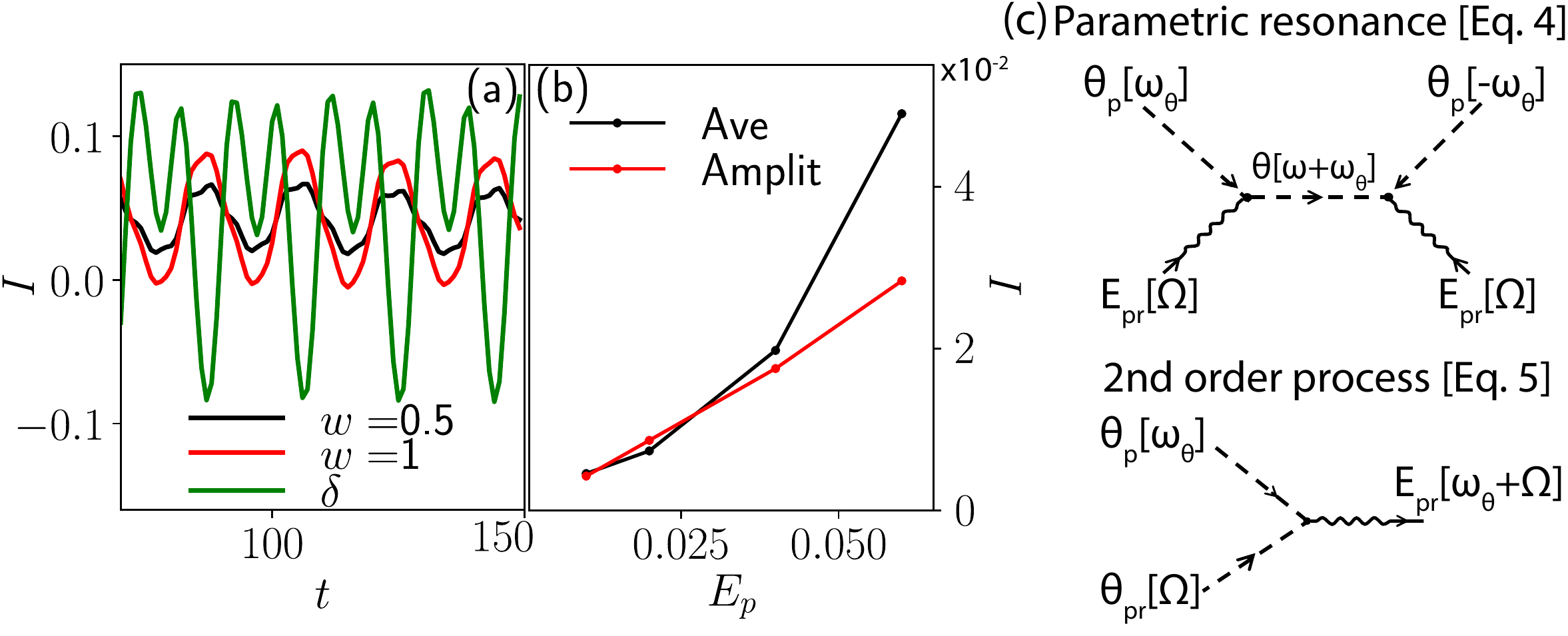}
\caption{(a)~Time evolution of the second harmonics' weight $I(t)$ for three different shapes of the probe pulse, namely $\delta$ is a short Gaussian pulse and a combination of two sinusoidal probes $w*\sin(\omega_{\theta})+\sin(2\omega_{\theta})$ with a long Gaussian envelope~(5 oscillations), where $w$ marks the relative weight. (b)~Scaling of the average value and the oscillation amplitudes in (a) versus the pump strength $E_0$ for $w=0.5$. (c)~Feynman diagram representation of the nonlinear optical response. The dashed lines mark the phase oscillations $\theta_p~(\theta_{\text{pr}})$ induced by the pump~(probe) pulse. The wiggly lines denote the electric field and the red dot is the vertex between the electric field and the phase. All results are for the ``primarily electronic'' case~(V=3.0,~$\lambda$=0.1,~$\Delta\epsilon=-2.4$, $\omega_0=0.1$).}
\label{Fig:scaling}
\end{figure} 

A complementary understanding can be obtained from the perspective of  nonlinear optics. The parametric resonance can be understood as the 3rd order optical process 
\begin{equation}
\frac{j(\omega)}{E_{\text{pr}}(\omega)}= \int d\tilde\omega_{ph}\sigma^{(3)}(\tilde\omega_{ph},-\tilde\omega_{ph},\omega) E_{p}(\tilde\omega_{ph}) E_{p}(-\tilde\omega_{ph}),
\label{Eq:Par}
\end{equation}
where $\sigma^{(3)}$ is the third-order current-current response function~\cite{sun2018third,parker2019} and $E_p~(E_{pr})$ marks the pump~(probe) pulse. However, the parametric resonance is not the only nonlinear response. In particular, there is a second order process that is generically expected to be the dominant contribution
\begin{equation}
\frac{j(\omega)}{E_{\text{pr}}(\omega)}= \int d\tilde\omega_{ph}\sigma^{(2)}(\tilde\omega_{ph},\omega-\tilde\omega_{ph}) E_{p}(\tilde\omega_{ph}) \frac{E_{pr}(\omega-\tilde\omega_{ph})}{E_{pr}(\omega)}.
\label{Eq:2nd}
\end{equation}

These processes are represented in terms of Feynman diagrams, see Fig.~\ref{Fig:scaling}(c).  In a realistic description, both processes will be present and we would like to separate their contributions. We study the difference on the example of the ``primary electronic'' driven insulator, see Fig.~\ref{Fig:phase}(a),\ by analyzing the weight of the photo-induced in-gap peak $I(t)=\int_{0.7}^{0.8}\sigma(\omega,t) d\omega$ as a function of time, see Fig.~\ref{Fig:scaling}(a). The time evolution of the weight $I$ show a positive average weight and persistent oscillations around the average. By considering different shapes of the probe pulse we exemplify that the average value is independent of the probe shape~(parametric resonance) while the oscillation amplitude depends on it~(second order). Moreover, the scaling with the pump amplitude $E_0$ shows quadratic~(linear) scaling for the average~(oscillator weight) value consistent with the parametric resonance~(2nd order) scenario, see Fig.~\ref{Fig:scaling}(b).

In conclusion, we have revealed that the time-dependent optical conductivity shows unique nonequilibrium signatures of collective behavior in the excitonic insulator. In particular, we have observed a new photo-induced in-gap state and attribute it to the nonlinear coupling of the phase mode distortion and the electromagnetic field leading to the \textit{parametric resonance}. On the other hand, the new in-gap state is suppressed for the ``primarily-lattice'' case.  The distinction allows us to identify the massive phase mode from the complicated in-gap optical response in excitonic insulator candidates~\cite{larkin2018,mor2018}. Its energy is a very sensitive measure of the relative electronic and lattice contribution to the gap opening and its experimental identification would provide important  insight into the nature of the gap opening. By evaluating the nonequilibrium optical conductivity, we propose that high-intensity teraherz pulses recently employed for the dynamical control of superconductivity~\cite{mitrano2016}, ferroelectricity~\cite{nova2019metastable,li2019terahertz} and nonlinear phononics~\cite{von2018probing,kozina2019terahertz} would represent a perfect setup for the parametric resonance of the phase mode and the corresponding dynamical phase transition.

\begin{acknowledgments}
The authors thank to T. Kaneko and M. Eckstein for fruitful discussions. The Flatiron Institute is a division of the Simons Foundation. YM acknowledges support from Grant-in-Aid for Scientific Re- search from JSPS, KAKENHI Grant Nos.~JP19K23425, JP20K14412, JP20H05265, and JST CREST Grant No.~JPMJCR1901. 
\end{acknowledgments}

%

\end{document}


\title{Supplemental materials for "Nonlinear response of collective modes in excitonic insulator"}
\author{Denis Gole\v{z}}
\affiliation{Center for Computational Quantum Physics, Flatiron Institute, 162 Fifth Avenue, New York, NY 10010, USA}
\author{Zhiyuan Sun}
\affiliation{Department of Physics, Columbia University, 538 West 120th Street, New York, NY 10027}
\author{Yuta Murakami}
\affiliation{Department of Physics, Tokyo Institute of Technology, Meguro, Tokyo 152-8551, Japan}
\author{Antoine Georges}
\affiliation{Center for Computational Quantum Physics, The Flatiron Institute, 162 5th Avenue, New York, NY 10010}
\affiliation{Department of Quantum Matter Physics, University of Geneva, 1211 Geneva 4, Switzerland}
\affiliation{CPHT, CNRS, Ecole Polytechnique, IP Paris, F-91128 Palaiseau, France}
\affiliation{Collège de France, 11 place Marcelin Berthelot, 75005 Paris, France}
\author{Andy Millis}
\affiliation{Department of Physics, Columbia University, 538 West 120th Street, New York, NY 10027}
\affiliation{Center for Computational Quantum Physics, The Flatiron Institute, 162 5th Avenue, New York, NY 10010}

\date{\today}

\maketitle
In these supplemental notes, we will first derive the low-energy effective action used in the main text at zero temperature, see Sec.~\ref{sec:gl_zero_T}. We will complement the derivation by the low-energy description close to the phase transition, see Sec.~\ref{Sec.:Action}. In Sec.~\ref{Sec:Optics}, we give details on the evaluation of the optical response and we decouple the optical response in the dipolar and the Peierls contribution in and out of equilibrium.

\section{Ginzburg-Landau action at zero temperature}
\label{sec:gl_zero_T}
\subsection{The action and the collective modes}
The action for the Hamiltonian in Eq.~(1) of the main text after the Hubbard-Stratonovich transformation reads
\begin{align}
\label{Eq:Seff}
S=\int dk d\tau \Psi^{\dagger}
\begin{pmatrix}
\partial_{\tau} + \epsilon_{k-A} & \phi + \sqrt{\lambda} X - ED \\
\bar{\phi} + \sqrt{\lambda} X - ED &
\partial_{\tau}  - \epsilon_{k-A}
\end{pmatrix} 
\Psi 
+ \frac{1}{U} |\phi|^2  + \frac{1}{2}X^2 + \frac{1}{\omega_0^2} \dot{X}^2
\end{align}
where $X$ is the generalized coordinate of the optical phonon at zero momentum, $\omega_0$ is its frequency, $\sqrt{\lambda}$ is the electron-phonon coupling strength and $\phi$ is the excitonic order parameter. For simplicity, we have introduced the relative energy $\epsilon_{k-A}=(\epsilon_{k-A,1}-\epsilon_{k-A,0})/2$ and  due to the particle-hole symmetry  the average energy vanishes $\ave{\epsilon_{k-A}}=(\epsilon_{k-A,1}+\epsilon_{k-A,0})/2=0.$ This phonon induces extra attractive interaction which enhances the excitonic gap. In the BCS limit, the free energy due to static uniform order parameter is
\begin{align}
F= \frac{1}{2}\nu |\phi + \sqrt{\lambda}X-ED|^2 \ln \frac{2\Lambda}{ |\phi + \sqrt{\lambda}X -ED|} + \frac{1}{U} |\phi|^2  + \frac{1}{2}X^2 
\label{eqn:free_energy}
\end{align} 
where $\nu$ is the normal state density of state and  $\Lambda\sim \varepsilon_F$ is the UV cutoff~(the Fermi energy). Assuming $\phi \in \mathcal{R}$ without loss of generality, variation with respect to uniform and static $\phi$ and $X$ gives the mean field gap equation
\begin{align}
\frac{\phi}{U}= \frac{X}{2\sqrt{\lambda}}  =\frac{1}{2} \sum_k \frac{\phi+\sqrt{\lambda} X}{E_k},
\end{align}
where $E_k=\sqrt{\epsilon_{k}^2 +  \Delta^2}$ is the renormalized quasiparticle energy and $\Delta=\phi+\sqrt{\lambda}X$ is the gap. Defining the enhanced effective coupling strength $U^\ast$, the gap equation simplifies to 
\begin{align}
\frac{1}{U^\ast} =\frac{1}{2} \sum_k \frac{1}{E_k}\,,
\quad
U^\ast \equiv U+U_X \equiv U+2\lambda
\end{align}
and the solution is $ \Delta= 2\Lambda e^{-\frac{2}{\nu U^\ast}-1/2}$ in the weakly coupling BCS regime. The excitonic order parameter is therefore $\phi=\Delta U/U^\ast$. 

Around the mean field minimum, the fluctuations of the order parameter and the phonon can be written as $\phi(r,t)=\phi+\delta(r,t)+i\phi \theta(r,t)$, where $\delta$, $\theta$ correspond to the amplitude and phase mode. For lattice degrees of freedom $X(r,t)= X+\sqrt{1/\nu}u(r,t)$ it is convenient to introduce the dimensionless phonon displacement $u$. The effective action for the fluctuations is 
\begin{align}
S[\delta,\theta, u] =  \sum_q \Bigg[ & \frac{1}{2}G^{-1}_{\theta}(q) \theta_{-q} \theta_{q} +\frac{1}{2}G^{-1}_{\delta}(q) \delta_{-q} \delta_{q} 
+\frac{1}{2}G^{-1}_{u}(q) u_{-q} u_{q} + \sqrt{\frac{\lambda}{\nu}}
\chi_{\sigma_1,\sigma_1}(q)  u_{-q} \delta_q 
 +
\chi_{\sigma_1,\sigma_2}(q) \theta_{-q}
\left( \delta_q + \sqrt{\frac{\lambda}{\nu}} u_{q} \right) 
\notag\\
& + \chi_{\sigma_1 \sigma_1}(q) D E_{-q} \left( -\frac{1}{2}\phi \theta^2_{q} +\delta_{q}
+\sqrt{\frac{\lambda}{\nu}} u_{q}\right) 
+\chi_{\sigma_2,\sigma_2, \sigma_1}(q,q_1) D E_{-q} \phi^2 \theta_{q_1} \theta_{q-q_1}
\notag\\
& + \chi_{\sigma_1 \sigma_1 \sigma_1}(q,q_1) D E_{-q}\sum_{q_1} (\delta + \sqrt{\frac{\lambda}{\nu}} u)_{q1} (\delta + \sqrt{\frac{\lambda}{\nu}} u)_{q-q1} 
\Bigg]\,,
\label{eqn:grand_action}
\end{align}
where the kernels for the three modes are
\begin{align}
& G^{-1}_{\theta}(q) =  \left(\frac{U}{U^\ast}\right)^2 K^{\mu\nu}(q)  q_\mu q_\nu
+ 2\left( \frac{1}{U} -\frac{1}{U^\ast} \right) \,,\quad
G^{-1}_{\delta}(q) =  \frac{2}{U} + \chi_{\sigma_1,\sigma_1}(q)
\,, \quad
G^{-1}_{u}(q) = \frac{\lambda}{\nu} \left(\frac{2}{U_X} + \chi_{\sigma_1,\sigma_1}(q) \right) -\frac{\omega^2}{\omega_0^2\nu}
\end{align}
and the correlation functions $\chi_{\sigma_i,\sigma_j}$ are defined in Ref.~\onlinecite{sun2020collective}. Note that $q$ marks both momentum and frequency in the above equations.

\subsubsection{Phase mode}
First, we analyze the action for the phase mode. The inverse propagator $G^{-1}_{\theta}(q)$ includes the kernel $K^{\mu\nu}(q)$  and without the electron-phonon coupling it is the same as for superconductors\cite{sun2020collective}. At zero momentum, the phase mode kernel simplifies to
\begin{align}
G^{-1}_{\theta}(\omega)
= \left[\chi_{\sigma_2,\sigma_2} + \frac{2}{U}
\right] \theta^2 
= \left[- \omega^2 F(\omega) + 2\left( \frac{1}{U} -\frac{1}{U^\ast} \right)
\right] \theta^2 
\end{align} 
where $F(\omega)=\sum_k \frac{1}{E_k(-\omega^2+4E_k^2)}$ contains the information of quasiparticle excitations \cite{sun2020collective}. The frequency of the phase mode is the root of the kernel and can be analytically analyzed in the BCS limit ($\Delta \ll \varepsilon_F$), where $F=\frac{\nu}{4\Delta^2} \frac{2\Delta}{\omega} \frac{\sin^{-1}\left( \frac{\omega}{2\Delta} \right)}{\sqrt{1-\left( \frac{\omega}{2\Delta} \right)^2}}$. We should comment that its linear coupling to the other modes has the kernel $\chi_{\sigma_1,\sigma_2}$ which is suppressed by the small number $\frac{\Delta}{\varepsilon_F}$. Thus we don't consider its influence on the phase mode frequency.

In the limit of weak electron-phonon coupling such that $U \approx U^\ast$, the phase mode develops a small gap $\omega_{\theta}=  2\Delta  \sqrt{\frac{1}{2} \left(\frac{1}{\nu U}- \frac{1}{\nu U^\ast}\right)}$. In the limit of strong electron-phonon coupling such that $U \ll U^\ast$, the phase mode frequency approaches the gap: $\omega_{\theta}=  2\Delta \left( 1- \frac{\pi^2}{32 d^2} (\nu U)^2 \right)$. This case corresponds to weak excitonic order and the phase mode is more like an s-exciton of the preformed semiconductor.
With fixed $U^\ast$, as $U$ increases from $0$ to $U^\ast$, corresponding to the weak and strong excitonic order, the phase mode frequency evolves from $2\Delta$ to $0$. This is similar to the Bardasis-Schrieffer mode  if one views $U$, $U^\ast$ as the coupling constants in the p-channel $\lambda_p$ and the s-channel $\lambda_s$ in Fig.~3 of Ref.~\onlinecite{sun2020BaSh}. 

\subsubsection{Amplitude mode and the phonon}
The amplitude mode kernel at zero momentum is 
\begin{align}
G^{-1}_{\delta}(\omega) =  \frac{2}{U} + \chi_{\sigma_1,\sigma_1}(\omega) =  \frac{2}{U} - \frac{2}{U^\ast} -(\omega^2 - 4\Delta^2) F(\omega)
\ge 0
\,.
\label{eqn:amplitude}
\end{align}
Without electron-phonon coupling, one has $U=U^\ast$ and \equa{eqn:amplitude} has a root at $\omega=2\Delta$ corresponding to the usual amplitude mode. With electron-phonon coupling $\lambda$, one has $\frac{2}{U} - \frac{2}{U^\ast} >0$ and $G^{-1}_{\delta}(\omega)  >0$, eliminating any root of \equa{eqn:amplitude}.

The phonon kernel is 
\begin{align}
G^{-1}_{u}(\omega) = \frac{\lambda}{\nu} \left(\frac{2}{U_X} + \chi_{\sigma_1,\sigma_1}(\omega) \right) -\frac{\omega^2}{\omega_0^2\nu}
=\frac{\lambda}{\nu}\left( \frac{2}{U_X} - \frac{2}{U^\ast} - \frac{\omega^2}{\lambda \omega_0^2}
+(4\Delta^2 -\omega^2) F(\omega)  \right) 
\,.
\label{eqn:phonon}
\end{align}
In general, formation of the gap pushes down the phonon frequency from $\omega_0$.
For $U_X < U_c$ where $U_c=\frac{1}{\frac{1}{U^\ast} +\frac{2\Delta^2}{\lambda \omega_0^2}}$, the dressed phonon frequency is above $2\Delta$. For $U_X > U_c$, the phonon is between $0$ and $2\Delta$.

Since the amplitude and phonon modes are coupled, one should look at the hybridized modes whose frequencies are given by the roots of the determinant: 
\begin{align}
det(M) &=
G^{-1}_{u}(\omega) G^{-1}_{\delta}(\omega) - \frac{\lambda}{\nu}  \chi_{\sigma_1,\sigma_1}^2 = \frac{\lambda}{\nu} \left( \frac{4}{U_X U} + (\frac{2}{U_X}+\frac{2}{U})\chi_{\sigma_1,\sigma_1}   \right) -\frac{\omega^2}{\omega_0^2 \nu} \left(\frac{2}{U} + \chi_{\sigma_1,\sigma_1}(\omega)\right)
\notag\\
&=\frac{\lambda}{\nu}\left( \frac{4}{U_X U} + (\frac{2}{U_X}+\frac{2}{U})\left(- \frac{2}{U^\ast} -(\omega^2 - 4\Delta^2) F(\omega)\right)  \right) -\frac{\omega^2}{\omega_0^2\nu} \left(\frac{2}{U} - \frac{2}{U^\ast} -(\omega^2 - 4\Delta^2) F(\omega)\right)
\notag\\
&=2\frac{\lambda}{\nu} \left[ 
-\frac{1}{U_X} \left(\frac{1}{U} - \frac{1}{U+U_X}\right) \frac{\omega^2}{\omega_0^2}\frac{\nu\omega_0^2}{\lambda}
+\left(
\frac{1}{U_X}+\frac{1}{U} -\frac{1}{U_X}\frac{\omega^2}{\omega_0^2}\frac{\nu\omega_0^2}{\lambda} \right)
(-\omega^2 + 4\Delta^2)F(\omega)
\right]
\,.
\label{eqn:amplitude_phonon}
\end{align}
From the behavior  $det(M)(\omega=0)=\frac{2\lambda}{\nu}(\frac{1}{U_X}+\frac{1}{U}) >0$ and $det(M)(\omega=2\Delta)=-\frac{4\Delta}{\lambda}\frac{1}{U(U+U_X)} <0$ and its detailed frequency dependence, one concludes that there is one and only one root below $2\Delta$, corresponding to the hybridized amplitude and phonon modes. If $\omega_0 < 2\Delta$, this mode is more like the softened phonon. If $\omega_0 > 2\Delta$, this mode is the amplitude mode pushed below $2\Delta$.

\subsection{Parametric resonance due to phase mode}
We assume that the electron-phonon coupling is weak such that the phase mode is well below the gap: $\omega_{\theta}\ll 2\Delta$.
In the low energy limit $\omega,v_F q \ll 2\Delta$, the kernel simplifies to $K^{\mu\nu}(q)=\text{diag}(-\nu, n/m)$.
Thus the low energy effective Lagrangian for the phase mode is 
\begin{align}
\mathcal{L}[\theta]= \left(\frac{U}{U^\ast}\right)^2 \left[-\frac{\nu}{2} \dot{\theta}^2 + \frac{n}{2m} (\nabla \theta)^2 + \frac{\nu}{2} \omega_{\theta}^2 \theta^2 
+ \frac{\nu}{2} \gamma_{\theta} E \theta^2 \right]
\end{align}
where $\omega_{\theta}=  \Delta  \sqrt{\frac{2}{\nu U} \left( 1- \frac{U}{U^\ast}\right)}$ is the electron-phonon interaction induced gap of the phase mode and $\gamma_{\theta}$ is the nonlinear coupling between the electric field $E$ and the phase $\theta$. Assuming the homogenous setup~($\nabla \theta=0$) we arrive to the equation of motion
\beq{
	\ddot \theta = -  [\omega_{\theta}^2 +  \gamma_{\theta} E] \theta.
	\label{Eq:EOM0}
}
The nonlinear coupling between $E$ and $\theta^2$ has two contributions:
\begin{align}
\frac{\nu}{2}\left(\frac{U}{U^\ast}\right)^2 \gamma_{\theta} &= -\frac{\nu}{4}\phi D \chi_{\sigma_1,\sigma_1}(0) - \frac{\nu}{4}\phi^2 D \chi_{\sigma_2,\sigma_2,\sigma_1}(0)
\notag\\
&= -\frac{\nu}{4}\phi D \sum_k \frac{-\epsilon^2}{E_k^3} - \frac{\nu}{4}\phi^2 D  \sum_k \frac{-{\Delta}}{E_k^3}
\notag\\
&\xrightarrow{BCS} \frac{\nu}{4}\phi D \left(-\nu+\frac{2}{U^\ast}\right) + \frac{\nu}{4} \phi^2 D \nu \frac{1}{{\Delta}}
=\frac{1}{2}\phi D \nu \left( \frac{\phi}{{\Delta}}-1 +\frac{2}{U^\ast \nu} \right)
\,,
\label{eqn:parametric_coupling}
\end{align}
see Appendix A1 of Ref.~\onlinecite{sun2020collective}. 
Another approach to obtain this coupling is to expand the static free energy in \equa{eqn:free_energy}:
\begin{align}
\left(\frac{U}{U^\ast}\right)^2 \frac{\nu}{2} \gamma_{\theta} E\theta^2 &= (\partial_{{\Delta}^2} F) D\phi E \theta^2 + \frac{1}{2} (\partial^2_{{\Delta}^2} F) 2{\Delta} (-2ED) \left[ \phi^2 \theta^2 -{\Delta} \phi \theta^2 \right] 
\notag\\
&= E \theta^2  D \phi  \left[ (\partial_{{\Delta}^2} F) -2  (\partial^2_{{\Delta}^2} F)  {\Delta} ( \phi  -{\Delta}  )
\right]
\notag\\
&= E \theta^2  D \phi  \left[ \sum_{k} \frac{1}{2E_k} +  \sum_{k} \frac{1}{2E_k^3}  \Delta ( \phi  -\Delta  )
\right]
\notag\\
&\xrightarrow{BCS} E \theta^2  D \phi  \left[ \frac{1}{U^\ast} + \frac{1}{2} \nu ( \frac{\phi}{\Delta}  -1  )
\right]
\,
\end{align}
which agrees with \equa{eqn:parametric_coupling}.
Therefore, 
\begin{align}
\gamma_{\theta}= D \Delta \left(\frac{U^\ast}{U}\right) \left(\frac{2}{\nu U^\ast} -1 +\frac{\phi}{\Delta}\right)
= D \Delta \left(\frac{U^\ast}{U}\right) \left(\frac{2}{\nu U^\ast} -1 +\frac{U}{U^\ast}\right)
\end{align}
in the weak coupling BCS case.

This coupling comes from the fact that fluctuation of the phase induces a dipole moment of $P=-\frac{\nu}{2}\left(\frac{U}{U^\ast}\right)^2 \gamma_{\theta}  \theta^2$  which couples to the electric field. Therefore, the electric field effectively modulates the frequency of the phase mode which induces parametric driving. If the electric field has a frequency of $2\omega_{\theta}$, this system forms a parametric oscillator and the phase  $\theta$ can absorb the energy from the drive, coined `parametric resonance'.

The power of dissipation scales with the amount of quantum or thermal fluctuations of the phase mode which are usually small. In the language of perturbation theory, this process is described by the correlator of dipole moment operators $\theta^2$, involving a loop formed by the phase mode propagator. However, if the phase mode is already coherently driven by some other source such that it has a large amplitude $\theta_0$ of oscillation, the parametric resonance is enhanced by $\theta_0^2.$ To show this, we assume that the system is oscillating due to some strong perturbation in the past and we will study the effect of a small and oscillating electric field $E(t)=E_0 e^{-\I \Omega t}$. We expand the phase motion as $\theta(t)=\theta_0 e^{\I \omega_{\theta} t}+\delta \theta(t).$ We insert this ansatz into Eq.~\ref{Eq:EOM0} and by neglecting terms of the order $\delta \theta E_0$ we arrive to
\beq{
 \delta \ddot \theta +\omega_P^2 \delta \theta = \gamma_{\theta} E_0 \theta_0 \exp^{\I (\omega_{\theta}-\Omega) t},
\label{eqn:parametric_resonance}
}
which is just a driven harmonic oscillator and the resonance is at $\Omega=2\omega_{\theta}.$ 

Now, we will evaluate the expression for the change in the optical conductivity. The linear response of $\theta$ to $E$ is 
\begin{align}
\theta^{(1)}(\omega+\omega_{\theta})= \frac{\gamma_{\theta} \theta_0(\omega_{\theta})}{\left( (\omega+\omega_{\theta})^2-\omega_{\theta}^2 \right)}  E(\omega)
\,
\end{align}
and the linear response current is thus 
\begin{align}
j^{(1)}(\omega)&=\partial_t P^{(1)}
=\left(\frac{U}{U^\ast}\right)^2 \frac{i\omega \gamma_{\theta}\nu}{2} \left( \theta^{(1)}(\omega+\omega_{\theta}) \theta_0(-\omega_{\theta}) +
\theta^{(1)}(\omega-\omega_{\theta}) \theta_0(\omega_{\theta}
)
\right)
\notag\\
&=\left(\frac{U}{U^\ast}\right)^2 \frac{i\omega \gamma_{\theta}\nu}{2} \left( \frac{\gamma_{\theta} \theta_0(\omega_{\theta})}{\left( (\omega+\omega_{\theta})^2-\omega_{\theta}^2 \right)}  E(\omega) \theta_0(-\omega_{\theta}) +
\frac{\gamma_{\theta} \theta_0(-\omega_{\theta})}{\left( (\omega-\omega_{\theta})^2-\omega_{\theta}^2 \right)}  E(\omega) \theta_0(\omega_{\theta}
)
\right)
\notag\\
&=\left(\frac{U}{U^\ast}\right)^2 \gamma_{\theta}^2\nu  
\frac{i\omega}{\omega^2-4\omega_{\theta}^2}   
\theta_0(\omega_{\theta})
\theta_0(-\omega_{\theta})
E(\omega)
\,.
\end{align}
The corresponding optical conductivity is
\begin{align}
\sigma_\theta=  \left(\frac{U}{U^\ast}\right)^2  \gamma_{\theta}^2\nu   \theta_0^2 \frac{i\omega}{\omega^2-4\omega_{\theta}^2}
=\nu D^2 \Delta^2 \left(\frac{2}{\nu U^\ast} -1 + \frac{U}{U^\ast} \right)^2    
\theta_0^2 \frac{i\omega}{\omega^2-4\omega_{\theta}^2}
\,
\label{eqn:phase_para_sigma}
\end{align}
which has a pole at twice the phason frequency. 

\subsection{Second order nonlinear optics due to the phase mode}
Light couples to the phase mode directly through the $\chi_{\sigma_2,\sigma_1}$ kernel. Moreover, light excites the amplitude/phonon mode linearly, which then excites the phase mode due to their linear coupling. In the small $\sqrt{\lambda}$ limit such that the electron-phonon coupling can be neglected, the effective linear response of the phase mode to the electric field is: 
\begin{align}
\theta^{(1)}(\omega)= \chi_{\theta E}(\omega) E(\omega) = G_\theta(\omega)\chi_{\sigma_1 \sigma_2}(\omega)
\left[1+
\tilde{G}_\delta(\omega)
\chi_{\sigma_1 \sigma_1}(\omega) 
+\tilde{G}_\delta(\omega) \chi_{\sigma_1 \sigma_2}(\omega) G_\theta(\omega)\chi_{\sigma_2 \sigma_1}(\omega)
\right]
D E(\omega) 
\,
\end{align}
where
\begin{align}
\tilde{G}_\delta(\omega)=\frac{G_\delta(\omega)}{1- \chi_{\sigma_1 \sigma_2}(\omega)  G_\delta(\omega) \chi_{\sigma_1 \sigma_2}(\omega) G_\theta(\omega)}
\,.
\end{align}
The second order current is thus 
\begin{align}
j^{(2)}(\omega)&=-\frac{\nu}{2}\left(\frac{U}{U^\ast}\right)^2 \gamma_{\theta}  \partial_t \left( \theta^2 \right) 
= \frac{\nu}{2}\left(\frac{U}{U^\ast}\right)^2 \gamma_{\theta} (-i\omega) \sum_{\omega_1+\omega_2=\omega}  \theta^{(1)}(\omega_1) \theta^{(1)}(\omega-\omega_1)
\notag\\
&=\frac{\nu}{2}\left(\frac{U}{U^\ast}\right)^2 \gamma_{\theta} (-i\omega) \sum_{\omega_1+\omega_2=\omega}  \chi_{\theta E}(\omega_1)  \chi_{\theta E}(\omega_2) E(\omega_1) E(\omega_2)
\,.
\end{align}

\subsection{Nonlinear optics due to the amplitude/phonon mode}
As shown by \equa{eqn:grand_action}, there is also nonlinear coupling between electric field and the amplitude/phonon mode. The nonlinear kernel $\chi_{\sigma_1,\sigma_1,\sigma_1}$ is a three point correction function, i.e., a one loop triangular diagram. At zero momentum, it can be nonzero if there are more than one quasiparticle bands, as in the case of excitonic insulators. Its static limit can be obtained by simply expanding the static free energy \equa{eqn:free_energy} which renders the nonlinear coupling term
\begin{align}
\mathcal{L}_{E,\delta,u}=\frac{\nu}{4\Delta } D  E (\delta+\sqrt{\lambda/\nu}u)^2
\,.
\end{align}
In the limit of $\omega \ll \Delta$, the phonon propagator \equa{eqn:phonon} becomes
\begin{align}
G^{-1}_{u}(\omega)
&=\frac{\lambda}{\nu}\left( \frac{2}{U_X} - \frac{2}{U^\ast} - \frac{\omega^2}{\lambda \omega_0^2}
+\nu(1 -\omega^2/4\Delta^2) \right) 
=\frac{1}{\nu}-\frac{\omega^2}{\omega_0^2\nu}-\frac{2\lambda}{\nu U^\ast}  +\lambda (1 -\omega^2/4\Delta^2)
\notag\\
&=-\left(1+\frac{\omega_0^2}{8\Delta^2} U_X \nu \right)\frac{\omega^2}{\omega_0^2\nu}+
\left(1-\frac{U_X}{U^\ast} + \frac{1}{2} U_X \nu \right)\frac{1}{\nu}
\,
\label{eqn:phonon_low_e}
\end{align}
which finally leads to the low energy effective Lagrangian
\begin{align}
\mathcal{L}[u]=-\frac{1}{2 \omega_0^2\nu}\left(1+\frac{\omega_0^2}{8\Delta^2} U_X \nu \right) \dot{u}^2+
\frac{1}{2}\left(1-\frac{U_X}{U^\ast} + \frac{1}{2} U_X \nu \right) \frac{u^2}{\nu} + \tilde \gamma_{\text{ph}} E  u^2
\,
\label{eqn:phonon_L}
\end{align}
where the nonlinear coupling between the electric field and the phonons is given by $\tilde \gamma_{\text{ph}}=\frac{\lambda D}{4\Delta}.$ Note that we did not write down the linear coupling terms  to electric field for notational simplicity. The optical conductivity due to parametric resonance is
\begin{align}
\sigma_u=4  \frac{\tilde\gamma_{\text{ph}}^2 \omega_0^2 \nu}{\left(1+\frac{\omega_0^2}{8\Delta^2} U_X \nu \right)}   u_0^2 \frac{i\omega}{\omega^2-4\omega_{u}^2}
=4  \frac{\gamma_{\text{ph}}^2 /(\omega_0^2 \nu)}{\left(1+\frac{\omega_0^2}{8\Delta^2} U_X \nu \right)}   u_0^2 \frac{i\omega}{\omega^2-4\omega_{u}^2}
,
\end{align}
where $\gamma_{\text{ph}}=\tilde \gamma_{\text{ph}} \omega_0^2\nu=\frac{D \nu\Delta}{4} \lambda \frac{\omega_0^2}{\Delta^2}.$ Therefore, both parametric resonance and second order nonlinear optical process can happen to the amplitude/phonon mode. Compared to \equa{eqn:phase_para_sigma} of  the phase mode, the spectra weight of $\sigma_u$ induced by parametric resonance with the phonon is smaller by the factor
\begin{align}
W_u/W_\theta
= \frac{  (\nu \lambda)^2 /2}{\left(1+\frac{\omega_0^2}{8\Delta^2} U_X \nu \right)  \left(\frac{2}{\nu U^\ast} -1 + \frac{U}{U^\ast} \right)^2}    
\left( \frac{\omega_0}{\Delta}\right)^4 
\left(\frac{u_0^2}{\omega_0^2 \nu^2 \theta_0^2} \right)
\end{align} 
The ratio is proportional to two small numbers, namely the dimensionless coupling $\nu \lambda$ and the frequency ratio $(\frac{\omega_0}{\Delta})^4$. In material candidates for the excitonic insulator phase, like 1T-TiSe$_2$ and Ta$_2$NiSe$_5$, the electron-phonon interaction can be substantial and $\nu \lambda$ can not be treated as a small number. In the main text, we have relaxed the weak coupling approximation by numerically solving the equation of motion. On contrary, in relavant materials the phonon frequency is much smaller than the gap size and the ration between the phonon frequencies and the gap size $(\frac{\omega_0}{\Delta})^4$ is a very small number. This explains why the nonlinear optical effects are srtongly suppressed for the phonon driven case. The physical intepretation of the ration is that the phonon degrees of freedom are heavy in contrast to the phase mode. Note that we have assumed that the pump induced phonon oscillation amplitude is related to that of phase by $u_0^2 \sim \omega_0^2 \nu^2\theta_0^2,$ which has been confirmed by a numerical simulation.

\section{Ginzburg-Landau action close to the transition temperature}\label{Sec.:Action}
Here, we will derive the Ginzburg-Landau action close to the transition temperature and show that the parametric resonance exist also close to the transition temperature. We will make an expansion of the Eq.~\ref{Eq:Seff} in the powers of the order parameter $\phi.$ It is convenient to defined the Green's function $G$ as
\bsplit{
G^{-1}&=\begin{pmatrix}
\partial_{\tau}  +\epsilon_{k-A,0} & \phi + \sqrt{\lambda} X -E D \\
\bar \phi + \sqrt{\lambda} X -E D & \partial_{\tau}  +\epsilon_{k-A,1} 
\end{pmatrix}=\begin{pmatrix}
\partial_{\tau} + \ave{\epsilon_{k-A}} - \epsilon_{k-A} & \phi + \sqrt{\lambda} X -E D  \\
\bar \phi + \sqrt{\lambda} X -E  D &  \partial_{\tau} +	  \ave{\epsilon_{k-A}} -\epsilon_{k-A} 
\end{pmatrix}
}
and we have used the average energy $\ave{\epsilon_{k-A}}=(\epsilon_{k-A,0}+\epsilon_{k-A,1})/2$ and the relative one $ \epsilon_{k-A}=(\epsilon_{k-A,0}-\epsilon_{k-A,1})/2.$ We will assume the particle-hole symmetric case and in this case $\ave{\epsilon_{k-A}}=0.$

\subsection{Mean field}
We can obtain the stationary point by stating $\delta S/\delta \phi_i=\delta S/\delta X_i=0$ and puting EM fields to zero. The derivative with respect to $ \phi_i$ gives the stationary value of the order parameter $\phi$:
\beq{
	\delta S/\delta \phi_i=0 \rightarrow \bar \phi/U= \sum_{k,n} \frac{ \bar \phi + \sqrt{\lambda}X}{\omega_n^2+\epsilon^2_k+(\bar \phi +\sqrt{\lambda}X)^2}= [\bar \phi + \sqrt{\lambda} X] \sum_k \frac{1-2 f(E_k)}{2 E_k},
}
where $E_k=\sqrt{\epsilon_k^2+\phi^2}$, $f(E)$ is the Fermi distribution function and the latter equation is just the usual gap equation. The variation with respect to the phonon distortion $X_i$ gives the stationary value of the distortion $X$
\beq{
	\delta S/\delta X_i=0 \rightarrow \omega_0^2 X = g Tr[G_{01}+G_{10}]= g \frac{\phi_0+\bar \phi_0}{U} \rightarrow X=\frac{2 g Re[\phi_0]}{\omega_0^2 U}. 
}
We can define the enhanced effective coupling strength $U^\ast=U+\frac{2g^2}{\omega_0^2} $ and arrive to the typical form of the gap equation
\begin{align}
\frac{1}{U^\ast} = \sum_k \frac{1-2f(E_k)}{2E_k}.
\end{align}

\subsection{Expansion close to the transition point}
We will separate the Greens function into the powers of the order parameter  and phonon distortions. The expansion will be made in real space and time. For latter convenience, we will introduce a generalized momentum operator $\hat k=(\hat k^0,\hat{k}) =(\I \partial_t,\vec \nabla)$ 
\bsplit{
G^{-1}=&G^{-1}_0 +X_1=\begin{pmatrix}
\hat k_0 -\epsilon_{\hat{k}-A} & 0  \\
0& \hat k_0 +\epsilon_{\hat{k}-A}\end{pmatrix} +
\begin{pmatrix}
0 & \phi + \sqrt{\lambda}X-E D   \\
\bar \phi + \sqrt{\lambda}X-E  D & 0 \end{pmatrix}
}

The effective actions is then given by
\bsplit{
	\I S_{eff}=& Tr[\log(G_0^{-1})] + Tr[\log(1-G_0 X_1)] + \frac{|\phi|^2}{U}= -Tr[\sum_{l} \frac{1}{2l}  (G_0 X_1)^{2l}] + \frac{ |\phi|^2}{U}
} 
and the quadratic term is given by
\bsplit{
	&\I S_{eff}^2= \frac{|\phi|^2}{U} - \tilde Tr[(\hat k^0- \epsilon_{\hat k-A})^{-1} X_1 (\hat k_0 + \epsilon_{\hat k-A})^{-1} X_1]  (\bar \phi+\sqrt{\lambda}X- E \bar D) (\phi+\sqrt{\lambda}X - E D),
}
where we have already performed the trace in the orbital space as marked by $\tilde Tr.$

We can use the derivative exansion~\cite{schakel2000time} which for a general functions $f(x)$ and $g(x)$ would be
\beq{
	f(x) \hat k_i g(x)= [\hat k_i -\I \partial_i] f(x)  g(x),
}
where $i$ goes through time and space indices. The advantage of this expansion is that $\hat k_i$ now can act on bra and reduces to a complex number. This means that we can rewrite the effective action as  
\bsplit{
	\I S_{eff}^2=& \frac{|\phi|^2}{U} - \tilde Tr[\frac{1}{k_0 +\epsilon_{k}} \frac{1}{k_0 - \partial_t + \epsilon_{k+\I\nabla} )} ] |\phi+\sqrt{\lambda}X-E D|^2 =
	 \frac{|\phi|^2}{U} - L(\partial_t,\vec \nabla) |\phi+\sqrt{\lambda}X-E D|^2,
}
where we have introduced the effective interaction between the pairing fields $L(\partial_t,\vec \nabla)$. We will now ask how to expand this term in the powers of the two derivatives. First, the trace represents the bubble diagram and the sum over frequencies can be performed
\beq{
L(\partial_t,\vec \nabla)= \frac{1}{N}\sum_{ k}  \frac{f(\epsilon_{k+\I \nabla}) - f(-\epsilon_{k})}{\I \partial_t  - [\epsilon_{k +\I \nabla} +\epsilon_{k}+\I \eta]}
}

In the equilibrium and for the homogenous case, the expression can be simplify to 
\beq{
	a= L_{\text{eq}}(\partial_t,\vec \nabla) = \frac{1}{N}\sum_{k}  \frac{\tanh(\epsilon_k/2T)}{\epsilon_k} \approx  \nu \int_0^{\Lambda} \frac{d \epsilon}{\epsilon} \tanh(\epsilon/2T) \propto \nu \ln(\Lambda /T),
}
where we have introduced the UV cut off $\Lambda.$ 

A similar expansion can be made for the quartic term and here we will neglect the spatial and temporal derivatives
\bsplit{
	&\I S_{eff}^4\approx  \tilde Tr\left[(\frac{1}{k_0^2- \epsilon_{k}})^{2}\right]  |\phi+\sqrt{\lambda}X-E D|^4=-\frac{1}{2}b  |\phi+\sqrt{\lambda}X-E D|^4
}
where close to the transition point $b=\nu c_f/T^2$ and $c_f$ is a constant of the order 1, see Ref.~\onlinecite{schakel2000time} for the detailed derivation.

Collecting the terms in the equilibrium and homogenous case we get
\beq{
	L_{\text{eff,eq}}= \frac{1}{U}  |\phi|^2 - a |\phi+\sqrt{\lambda}X-E D|^2 -\frac{b}{2}|\phi+\sqrt{\lambda} X-E D|^4
}
and separating the absolute and the phase value of the order parameter $\phi=|\phi| \exp^{\I \theta}$ we get
\beq{
	L_{\text{eff},\text{eq}}= \left(\frac{1}{U}-a\right) |\phi|^2 +2 (\sqrt{\lambda}X-E D) |\phi| a \cos(\theta) - b/2~\left[\phi^2+ 2 (\sqrt{\lambda}X-E D) |\phi| a \cos(\theta) + \lambda X^2 \right]^2 + \lambda X^2.
}
Using the mean-field value for the order parameter, we get for the harmonic part of the phase mode potential $\sqrt{\lambda} X\phi a.$ To determine the gap of the phase mode we also need to evaluate the effective mass.

\subsection{Derivative expansion in time}
Now, we do expansion in the time derivative
\bsplit{
	L(\partial_t,0)=&\frac{1}{N}\sum_k \frac{f(\epsilon_k)-f(-\epsilon_k)}{k_0 - 2\epsilon_k +\I \eta}=\nu \int d\epsilon \frac{1-2f(\epsilon)}{k_0 - 2\epsilon +\I \eta}=\nu \mathcal{P} [\int d\epsilon \frac{1-2f(\epsilon)}{k_0 - 2\epsilon }] - \I \nu  2 \pi \int d\epsilon [1-2f(\epsilon/2)] \delta (k_0-2 \epsilon),
}
where the $\mathcal{P}$ marks the principal value. By xpanding the imaginary part into the Taylor series in $k_0$ we get
\beq{
	\text{Im[}L(\partial_t,0)]\approx-\nu \frac{\pi}{2} \beta k_0 + O(k_0^3), 
}
where we have introduced the constant $c=\nu \frac{\pi}{2} \beta.$
The real part of the expression can be simplified to
\beq{
	\text{Re[}L(\partial_t,0)]\approx-\nu P \left[\int d\epsilon \frac{1-2f(\epsilon)}{-2\epsilon } ( 1+\frac{k_0}{2\epsilon}-\frac{k_0^2}{4\epsilon^2})\right] + O(k_0^3) = a + d k_0^2  + O(k_0^3).
}
In the BCS regime, the constant  $d$ reduces to  $d=3\xi_0^2/v_F^2$ and we have introduced the BCS correlation lenght $\xi_0^2=\frac{7\zeta(3)}{48\pi^2}\frac{v_F^2}{T_c^2}$ and $T_c$ is the transition temperature  and the $v_F$ is the Fermi velocity. 

Now, we can express the dynamical part of the effective action as 
\bsplit{
	L_{\text{eff,dyn}}= c (\phi^*+\sqrt{\lambda}X-ED)[ (\I \partial_t)] (\phi+\sqrt{\lambda}X-ED)+ \\ 	
	 d (\phi^*+\sqrt{\lambda}X-ED)[ (\I \partial_t)^2] (\phi+\sqrt{\lambda}X-ED)
}
In this case, the separation on the absolute and the phase directions gives additional terms 
\bsplit{
	&L_{\text{eff,dyn}}= \\
	&\I c \left( |\phi| \dot{|\phi|} +\I \dot \theta |\phi|^2 + |\phi| \exp^{-\I \theta} [\sqrt{\lambda}\dot X +\dot E D] + \dot{|\phi|} \exp^{\I \theta} [\sqrt{\lambda}X + E D] +  |\phi| \exp^{\I \theta} [\I\dot \theta] [\sqrt{\lambda}X + E D]  + (\sqrt{\lambda}X+E D) (\sqrt{\lambda} \dot X +\dot E D) \right)  \\
	&d\left( \dot{|\phi|}^2 + |\phi|^2 \dot \theta^2 + \lambda \dot X^2 -\I 2 |\phi| \sqrt{\lambda} \dot X \dot \theta \cos(\theta) + 2 \sqrt{\lambda}\dot \phi \dot X \cos(\theta) - 2 \I \phi \dot \phi  \dot \theta - 2 D \left[ \dot E \dot \phi \cos(2\theta) + \sqrt{\lambda} \dot X \dot E -\I \dot E \phi \dot \theta + D \dot E^2 \right] \right)
}
From the linear term, we can see that the electric field couples linearly to the phase mode as seen in optics. 

\paragraph{Phase mode and phonons}
In this part, we would like to understand only the coupled dynamics of the phase mode and phonon. Therefore we neglect the variation of the amplitude mode $\dot \phi=0.$ The effective Lagrangian in this case can be written as
\bsplit{
	L_{\text{eff}}&= \left(-\frac{1}{U}+a\right) |\phi|^2 +2 (\sqrt{\lambda}X-ED) |\phi| a \cos(\theta) + \lambda X^2  - b/2~[\phi^2+ 2 (\sqrt{\lambda}X-E D) |\phi| a \cos(\theta) + \lambda X^2 ]^2 \\
	&+ d\left[ |\phi|^2 \dot \theta^2 - 2 \I \sqrt{\lambda} |\phi| \dot \theta \dot X \cos(\theta) -\lambda \dot X^2 +\dot E^2 D^2  -2\sqrt{\lambda}D \dot X \dot E -2\I E D |\phi| \dot \theta \cos(\theta)  \right] + \\
	&+ \I c\left[ \I \dot \theta |\phi| + |\phi| \exp^{\I \theta} (\sqrt{\lambda} \dot X+\dot E D) + |\phi| \exp^{\I \theta} \I \dot \theta (\sqrt{\lambda X}+ED) +(\sqrt{\lambda}X+ED)(\sqrt{\lambda}\dot X+\dot ED)  \right] 
}
The terms proportional to the $c$ will introduce damping into the system and only exist close to the transition temperature. In the following, we will neglect its effect by putting $c\rightarrow0.$ We can now derive the equations of  motion for the phase direction
\bsplit{
	\partial L_{\text{eff}}/\partial \theta=& -a 2 (\sqrt{\lambda}X-ED) |\phi|  \sin(\theta) +2  \I \sqrt{\lambda} d |\phi| \dot \theta \dot X \sin(\theta) +2 \I E D d |\phi| \dot \theta \sin(\theta ) + \\
	&b [\phi^2+ 2 (\sqrt{\lambda}X-E D) |\phi| a \cos(\theta) + \lambda X^2 ] 2(\sqrt{\lambda} X -E D) |\phi| a \sin(\theta)  \\ 
	d(\partial L_{\text{eff}}/\partial\dot \theta)/dt&=2 d |\phi|^2  \ddot \theta +2 \I \sqrt{\lambda} |\phi|  d \dot X  \sin(\theta) \dot \theta +2 \I ED d |\phi| \sin(\theta)\dot \theta
}
Assuming the static phonon distortion $X$ we arrive to the equation of motion for the phase mode
\beq{
	\ddot \theta = - \frac{\sqrt{\lambda}X a}{d|\phi|} \sin(\theta) + \frac{a E D}{d|\phi|}  \sin(\theta) + O(b a ),
	\label{Eq:EOM}
}
where we have neglected the corrections proportional to the small parameter $b$. 

Without the electric field $E=0$ and by expanding in the small phase distortion $\sin(\theta)\approx \theta$, we observe that the effective action is equivalent to the harmonic oscillation
\beq{
	\ddot \theta = - \omega_{\theta}^2 \sin(\theta) + O(b a ),
}
and the gap of the phase mode is given by $\omega^2_{\theta}=\frac{\sqrt{\lambda}X a}{d|\phi|}.$

\section{Optical conductivity and effect of the dipolar moment}\label{Sec:Optics}

For completeness of the paper, we will first describe how we evaluate the time-resolved optical conductivity. Next, we introduce the linear response formalism in equilibrium for the excitonic order and optical conductivity, see Ref.~\onlinecite{murakami2020} for details. We will use these  expressions for a simple extension to the static but nonthermal states.  Then we will present which features in the optical conductivity originate from the dipolar and Peierls contribution for the equilibrium case and after the photo-excitations.

The photo-induced current is given by a time-derivative of the polarization $ j= j_{\text{intra}}+ j_{\text{inter}}=\partial (P_{\text{intra}}+ P_{\text{inter}})/\partial t$ and we have separate the intraband $ j_{\text{intra}}$ and the interband $ j_{\text{inter}}$ contributions. The intra-band contribution follows from the usual Peierls expression $j_{\text{inter}}=q \sum_{k\alpha} v_{k\alpha} c_{k\alpha}^{\dagger} c_{k\alpha},$ where we have introduce the effective velocity $v_{k\alpha}=\partial \epsilon_{k\alpha}/\partial_k.$ The interband current can be obtained from the derivative of the polarization $\vec j_{\text{intra}}=\frac{\partial P_{\text{inter}}}{\partial t}$ and in the following we will evaluate it numerically. We will analyze the optical absorption by explicit simulations of the pump and probe pulse and extract the photo-induced current as the difference of the current with and without a probe pulse $j_{\text{probe}}=j_{\text{probe+pump}}-j_{\text{pump}}.$ We will use the Fourier transform $B(\omega,t_p)=\int_0^{t_{\text{cut}}} ds B(t_p + s) e^{-\I \omega s-\eta s}$ for  $X = j_{\text{probe}}$ or electric field $E$, where $t_p$ is the time of the probe pulse and we will use a small broadening $\eta = 0.03$ to avoid the artifacts originating from the finite Fourier window.  Finally, the optical conductivity is evaluated as a derivative of the photo-induced current and the probe field \beq{\sigma(\omega,t_p)=\frac{j_{\text{probe}}(\omega,t_p)}{E_{\text{probe}}(\omega,t_p)}.}  

In equilibrium, the excitonic response function is given by
\beq{
	\chi_{\mu\nu}(t-t',i-j)=-\I\theta(t-t')\ave{[\rho_{\mu i},\rho_{\nu j}]},
}
where we have parametrized the local density matrix as $\rho_{\mu,i}=\frac{1}{N}\sum_i \psi_i^{\dagger} \sigma_{\mu} \psi_i,$ where $\sigma_{\mu}$ are the Pauli matrices for $\mu=0,1,2,3.$ In equilibrium and for translational invariant system, we can introduce a Fourier transform $\chi_{\mu\nu}(\omega,q)=\sum_i \int dt \chi_{\mu\nu}(t,i) e^{\I\omega t-\I q i}.$ 

Within the random phase approximation, the excitonic response function is obtained as
\beq{
	\chi_{\mu\nu}(\omega,q)=\chi_{\mu\nu}^0(\omega,q) + \chi_{\mu\nu}^0(\omega,q) \theta(\omega,q) \chi_{\mu\nu}(\omega,q),
}
where for the Hamiltonian in Eq.~1~(main text) the vertex expresses as $\theta(\omega,q)=\text{Diag}[V/2,-V/2+\lambda D_0(\omega),-V/2,-V/2]$ and $\text{Diag}$ marks the diagonal matrix. We have introduced a free phonon Green's function $D_0(\omega)=\frac{2\omega_0}{\omega^2-\omega_0^2}.$  The  bare bubble contribution $\chi_{\mu\nu}^0(\omega,q)$ is expressed as a generalized Linhard-like formula
\bsplit{
\chi_{\mu\nu}^0(\omega;q)=\frac{1}{N}\sum_{\bf k} \sum_{a,b=\pm}{\rm Tr}[W_a(k-q) \sigma_{\mu} W_b(k) \sigma_{\nu}]\nonumber \times\frac{f(E_a({k-q}))-f(E_b({k}))}{\omega-(E_b({k})-E_a({k-q}))},
}
where we have introduced the Bogoliubov quasiparticle energy $E_{\pm}(k)=\pm\sqrt{\epsilon_k^2+|\Delta+\sqrt{\lambda}X|^2}$  and the corresponding eigenvectors $W_{\pm}(k).$ The bare bubble $\chi^0$ corresponds to the non-self-consistent evaluation of the time-dependent mean-field equation with a small source field, while the full random phase approximation $\chi$ is equivalent to the self-consistent mean-field equations. 

Now we explain how is the excitonic response function connected with the optical response. By expanding the kinetic term in the vector potential we obtain
\beq{
	H=H_0  + \vec j_{\text{inter}} \vec A + \vec P_{\text{intra}} \vec E= \sum_{\nu=1}^3 O_{\nu} \sigma_{\nu} F_{\nu}^{ext},
}
where we have introduced Pauli matrix $\sigma_{\nu},$ corresponding operators $O_{\nu}=[\vec P_{\text{intra}},0,\vec j]$ and fields $F_{\nu}^{ext}=(\vec E,0,\vec A).$ We are interested in a linear conductivity for induced polarizations and currents 
\bsplit{
	\vec O^{\nu}(t)=&\int dt' \chi^{\nu\mu}(t,t') F_{\mu}^{ext}(t').
}
Here, we need to connect the susceptibility $\chi_{\nu\mu}$ with the optical conductivity $\sigma=\sum_{\nu} \sigma_{\nu}.$ First, we note that the photo-induced current due to the change of the polarization is $\vec j_{\text{intra}}=d \vec P_{\text{intra}}/dt=D d \vec \rho_{1,\text{loc}}/dt $ and the vector potential is connected with the electric field as $\vec E=-d\vec A/dt.$ The optical conductivity can be separated into the dipolar contribution $\sigma_{1}(\omega)=\green{j_{\text{intra}}}{P_{\text{intra}}}(\omega)$ and the Peierls contribution $\sigma_{3}(\omega)= \green{j_{\text{inter}}}{j_{\text{inter}}}(\omega),$ while the mixed intra- and inter-band components are zero due to the inversion symmetry. 
In equilibrium, the dipolar contribution is expressed as $\sigma_{1}(\omega)=  \green{j_{\text{intra}}}{P_{\text{intra}}}(\omega)=D^2 \I \omega \chi_{11}(\omega)$ and the Peierls contribution as $\sigma_{3}(\omega)= \green{j_{\text{inter}}}{j_{\text{inter}}}(\omega)= \chi_{33}(\omega)/(\I \omega)$.

\begin{figure}[t]
\includegraphics[width=0.9\linewidth]{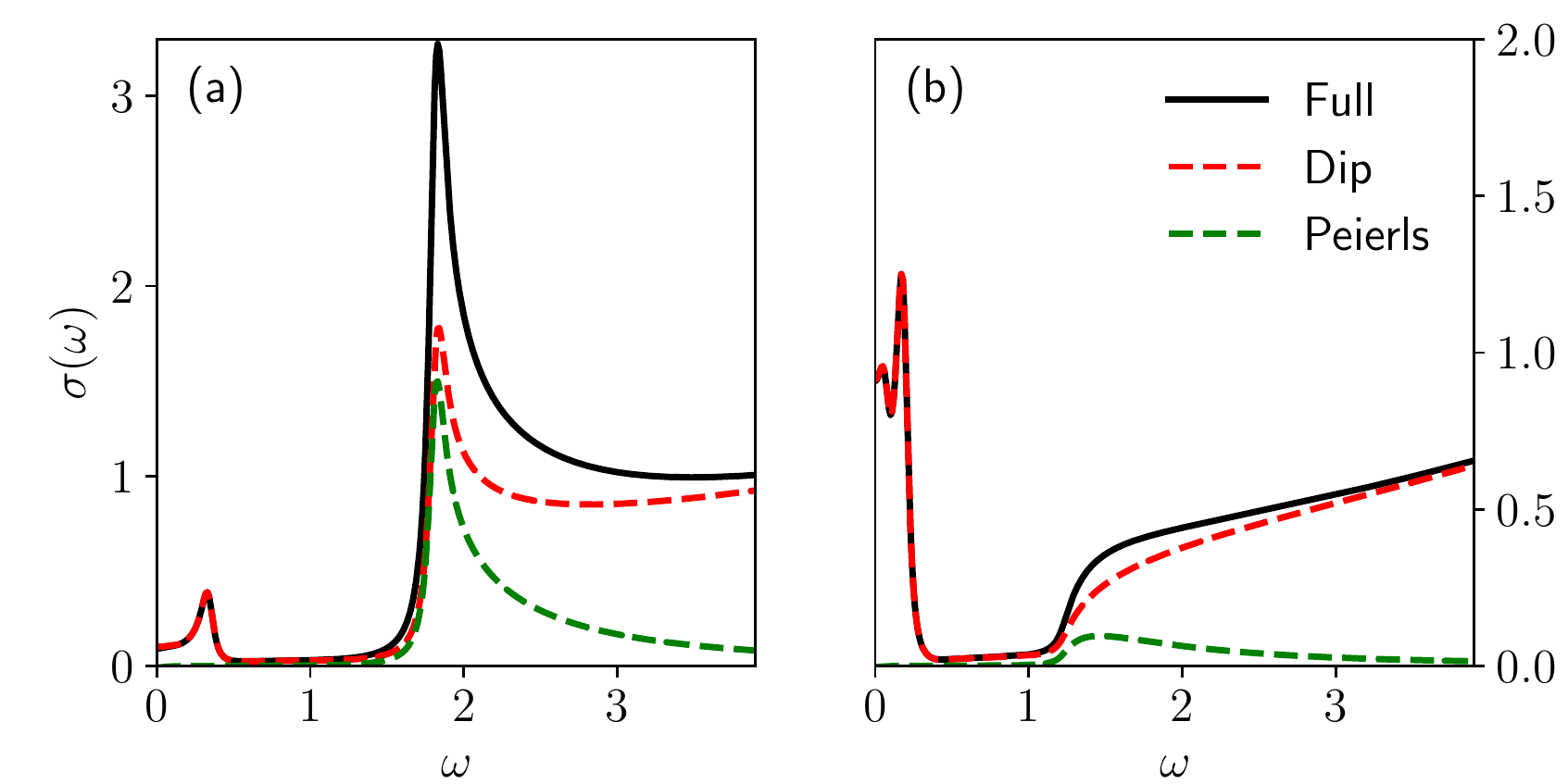}
\caption{Decoupling of the equilibrium optical conductivity $\sigma(\omega)$ into the dipolar contribution $\sigma_1(\omega)$ and the Peierls contribution $\sigma_3(\omega)$ for the BCS~(a) and the BEC regime~(b) at zero temperature. For the BCS case, the parameters are the same as in the main text. For the BEC case, the crystal-field splitting is $\Delta_1=-\Delta_0=2.4$, the phonon frequency $\omega_0=0.1$ and the el.-ph. interaction $\lambda=0.28$ }
\label{Fig:optics_decoupled}
\end{figure}

In Fig.~\ref{Fig:optics_decoupled}, we analyze the contribution of the dipolar and the Peierls contribution to the optical conductivity in equilibrium for the same example as in the main text. In the BCS regime, the peak at the gap edge shows a similar contribution from the Peierls and the dipolar term. On the contrary, the in-gap peak corresponding to the massive phase mode is of purely dipolar origin. Its weight is proportional to the square of the dipolar matrix element $D$ as seen from the relation between the dipolar part of the optical conductivity and the susceptibility of the order parameter $\sigma_{1}(\omega)=D^2 \I \omega \chi_{11}(\omega).$ A similar observation is valid for the BEC case, except that most of the contribution at the gap edge originates from the dipolar term, see Fig.~\ref{Fig:optics_decoupled}(b). 

%
